\documentclass[twocolumn,showpacs,floats,floatfix,superscriptaddress,aps,pra]{revtex4-1}
\usepackage{amsfonts}
\usepackage{amssymb}
\usepackage{amsmath}
\usepackage{calc}
\usepackage{graphicx}
\usepackage{bm}

\usepackage[normalem]{ulem}
\usepackage{amsmath,amssymb}
\usepackage{multirow}
\usepackage{xcolor,soul}
\usepackage{srcltx}
\usepackage{hyperref,graphicx}

\def\be{ \begin{equation}}
\def\ee{ \end{equation}}
\def\bea{ \begin{eqnarray}}
\def\eea{ \end{eqnarray}}
\def\bse{ \begin{subequations}}
\def\ese{ \end{subequations}}
\def\bc{ \begin{center}}
\def\ec{ \end{center}}

\begin{document}

\author{Stefano Longhi}
\email{stefano.longhi@.polimi.it}
\affiliation{Dipartimento di Fisica, Politecnico di Milano and Istituto di Fotonica e Nanotecnologie del Consiglio Nazionale delle Ricerche, Piazza L. da Vinci 32, I-20133 Milano, Italy}

\title{Convective and absolute $\mathcal{PT}$ symmetry breaking in tight-binding lattices}
\date{\today }

\begin{abstract}
We investigate  the onset of parity-time ($\mathcal{PT}$) symmetry breaking in non-Hermitian tight-binding lattices with spatially-extended loss/gain regions in presence of an advective term. Similarly to the instability properties of hydrodynamic open flows, it is shown that $\mathcal{PT}$ symmetry breaking can be either absolute or convective. In the former case, an initially-localized wave packet  shows a secular growth with time at any given spatial position, whereas in the latter case 
the growth is observed in a reference frame moving at some drift velocity while decay occurs at any fixed spatial position. In the convective unstable regime, $\mathcal{PT}$ symmetry is restored when the spatial region of gain/loss in the lattice is  limited (rather than extended). We consider specifically a non-Hermitian extension of the Rice-Mele tight binding lattice model, and show the existence of a transition from absolute to convective symmetry breaking when the advective term is large enough. 
An extension of the analysis to ac-dc-driven  lattices is also presented, and an optical implementation of the non-Hermitian Rice-Mele model is suggested, which is based on light transport in an array of evanescently-coupled optical waveguides with a periodically-bent axis and alternating regions of optical gain and loss.
\end{abstract}

\pacs{
03.65.-w, 
11.30.Er, 
72.10.Bg, 
42.82.Et, 
}
\maketitle

\section{Introduction}
Non-Hermitian Hamiltonian models are often encountered in a wide class of quantum and classical systems \cite{Moiseyev}. They are introduced, for example, to model open systems and dissipative phenomena in quantum mechanics  (see, for instance, \cite{Moiseyev,Rotter,ob1,ob2,vi,kor}). In optics, non-Hermitian models naturally arise owing to the presence of optical gain and loss regions in dielectric or metal-dielectric structures \cite{Siegman}. A special class of non-Hermitian Hamiltonians is provided by complex potentials having parity-time ($\mathcal{PT}$) symmetry \cite{Bender_PRL_98,Bender_RPP_2007}, that is invariance under simultaneous parity transform ($\mathcal{P}$: $\hat{p} \rightarrow -\hat{p}$, $\hat{x} \rightarrow -\hat{x}$, where $\hat{p}$ and $\hat{x}$ stand for momentum and position operators, respectively) and time reversal ($\mathcal{T}$: $\hat{p} \rightarrow -\hat{p}$, $\hat{x} \rightarrow \hat{x}$, $i\rightarrow-i$). An important property of $\mathcal{PT}$ Hamiltonians is to admit of an entirely real-valued energy spectrum below a phase transition symmetry-breaking point, a property that attracted great attention in earlier studies on the subject owing to the possibility to formulate a consistent quantum mechanical
theory in a non-Hermitian framework \cite{Bender_PRL_98,Bender_RPP_2007,Mos1,Ben}.  Indeed, 
$\mathcal{PT}$-symmetric Hamiltonians are a
special case of pseudo-Hermitian Hamiltonians, which can be mapped into Hermitian ones \cite{Mos2}.
$\mathcal{PT}$-symmetric  Hamiltonians have found interest and applications in several physical fields, including magnetohydrodynamics \cite{Guenther_JMP_2005}, cavity quantum electrodynamics \cite{Plenio_RMP_1998}, quantum-field-theories \cite{Ben,BenQFT}, and electronics \cite{Kottos}. More recently, great efforts have been devoted to the study  and the experimental implementation   of optical structures possessing $\mathcal{PT}$ symmetry (see, for instance, \cite{Muga,El-Ganainy_OL_07, Makris_PRL_08, Klaiman_PRL_08, Mostafazadeh_PRL_09, Longhi_PRL_09, Guo09,Ruter_NP_10, Longhi10,Feng2011,Kivshar12,Regensburger_Nature_12,Feng12,uffa} and references therein). The huge interest raised by the introduction of $\mathcal{PT}$ optical media is mainly motivated by their rather unique properties to mold the flow of light in non-conventional ways, with the possibility to observe, for example,   double refraction and nonreciprocal diffraction patterns \cite{Makris_PRL_08}, unidirectional Bragg scattering and invisibility \cite{Longhi_PRL_09, Lin_PRL_2011, Regensburger_Nature_12, LonghiPRA10, Longhi_JPA_2011,Graefe11,Feng12,MosIN}, non-reciprocity \cite{nonlinearPT},
giant Goos-H\"anchen shift \cite{Longhi_PRA_2011}, and simultaneous perfect absorption and laser behaviour \cite{Longhi_PRA_2010, Chong_PRL_2011}. So far, $\mathcal{PT}$ quantum and classical systems have been mainly investigated in the unbroken $\mathcal{PT}$ phase, where the energies are real-valued, or at the symmetry breaking point, where exceptional points or spectral singularities appear in the underlying Hamiltonian (see, for instance, \cite{Klaiman_PRL_08,Longhi10,MosRes}).  In the broken $\mathcal{PT}$ phase, 
complex-conjugate energies appear. In the context of spatially-extended dissipative dynamical systems and hydrodynamic flows \cite{Cross}, breaking of the $\mathcal{PT}$ phase indicates a bifurcation from a marginally-stable phase to an unstable phase.  
This means that, while an initially localized wave packet can not secularly grow in the unbroken $\mathcal{PT}$ phase, it does in the broken $\mathcal{PT}$ phase owing to the emergence of modes with complex energies. In hydrodynamics, an unstable open flow can be classified as either {\it absolutely} or {\it convectively} unstable \cite{rev1,flow1,flow2}.  A one-dimensional flow described by an  order parameter $\psi(x,t)$ is unstable if, for any given localized perturbation $\psi(x,0)$ at initial time $t=0$, $\psi(x,t) \rightarrow \infty$ as $t \rightarrow \infty$ along at least one ray $x/t=v={\rm const}$. The instability is said to be absolute if $\psi(x,t) \rightarrow \infty$ along the ray $x/t=0$,  whereas it is convective if $\psi(x,t) \rightarrow 0$ along the ray $x/t=0$ \cite{rev1}. Physically, in the convectively unstable regime the initial perturbation grows when observed along the trajectory $x=vt$ at some drift velocity $v$, whereas it  decays when observed at a fixed position.  Convectively unstable flows generally arise in the presence of an advective (drift) term in the system, in such a way that the growing perturbation drifts in the laboratory reference frame and eventually escapes from the system. Originally introduced in hydrodynamic contexts, 
the concepts of convective and absolute instabilities have found interest and applications in other physical fields, for example in the study of dissipative optical patterns and noise-sustained structures in nonlinear optics \cite{Santa}.
\par 
Inspired by the properties of hydrodynamic unstable flows \cite{rev1}, in this work we introduce the concepts of convective and absolute $\mathcal{PT}$ symmetry breaking for spatially-extended Hamiltonian systems.  Specifically, we investigate the symmetry-breaking properties of a tight-binding lattice model with spatially-extended alternating gain and loss regions, and show that the presence of an advective term can change the symmetry breaking from absolute to convective. The lattice model that we consider is a non-Hermitian extension of the famous Rice-Mele Hamiltonian, originally introduced to model conjugated diatomic polymers \cite{Rice}. In the convectively $\mathcal{PT}$ symmetry breaking regime, the $\mathcal{PT}$ symmetry can be restored when the gain/loss region becomes spatially confined. A physical implementation of the non-Hermitian Rice-Mele lattice model is proposed using  arrays of coupled optical waveguides in a zig-zag geometry with periodically-bent axis and alternating optical gain and loss.\par
The paper is organized as follows. In Sec.II the Rice-Mele tight-binding lattice model with non-Hermitian and advective terms is presented, and a physical implementation based on light transport in arrays of coupled optical waveguides is suggested. In Sec.III the concepts of absolute and convective $\mathcal{PT}$ symmetry breaking are introduced for periodic potentials, and the transition from absolute to convective symmetry breaking for the Rice-Mele lattice model is studied by application of asymptotic (saddle point) methods. The concepts of convective and absolute symmetry breaking are also discussed for ac-dc driven lattice models, where the quasi-energy bands . In Sec.IV the main conclusions and future developments are outlined. Finally, in two Appendixes some technical details on Floquet analysis of the ac-dc driven lattice model and saddle point calculations for the Rice-Mele Hamiltonian are presented.
\section{The model}
\subsection{Extended Rice-Mele Hamiltonian and $\mathcal{PT}$ symmetry breaking}
We consider transport of classical or quantum waves on a $\mathcal{PT}$-invariant  tight-binding dimerized superlattice with nearest and next-nearest neighborhood hopping schematically shown in Fig.1(a).  The  evolution of the amplitude probabilities $a_n(t)$, $b_n(t)$ at the two sites of the $n$-th unit cell in the lattice is governed by the following coupled-mode equations
\begin{eqnarray}
i \frac{d a_n}{d t} & = & -\kappa b_n-\sigma b_{n-1}-\rho \exp(i \varphi) a_{n+1} \nonumber \\
 & - & \rho \exp(-i \varphi) a_{n-1}+ig a_n \\
i \frac{d b_n}{d t} & = & -\kappa a_n-\sigma a_{n+1}-\rho \exp(i \varphi) b_{n+1} \nonumber \\
& - & \rho \exp(-i \varphi) b_{n-1}-ig b_n 
\end{eqnarray}

\begin{figure}[t]
\includegraphics[width=8cm]{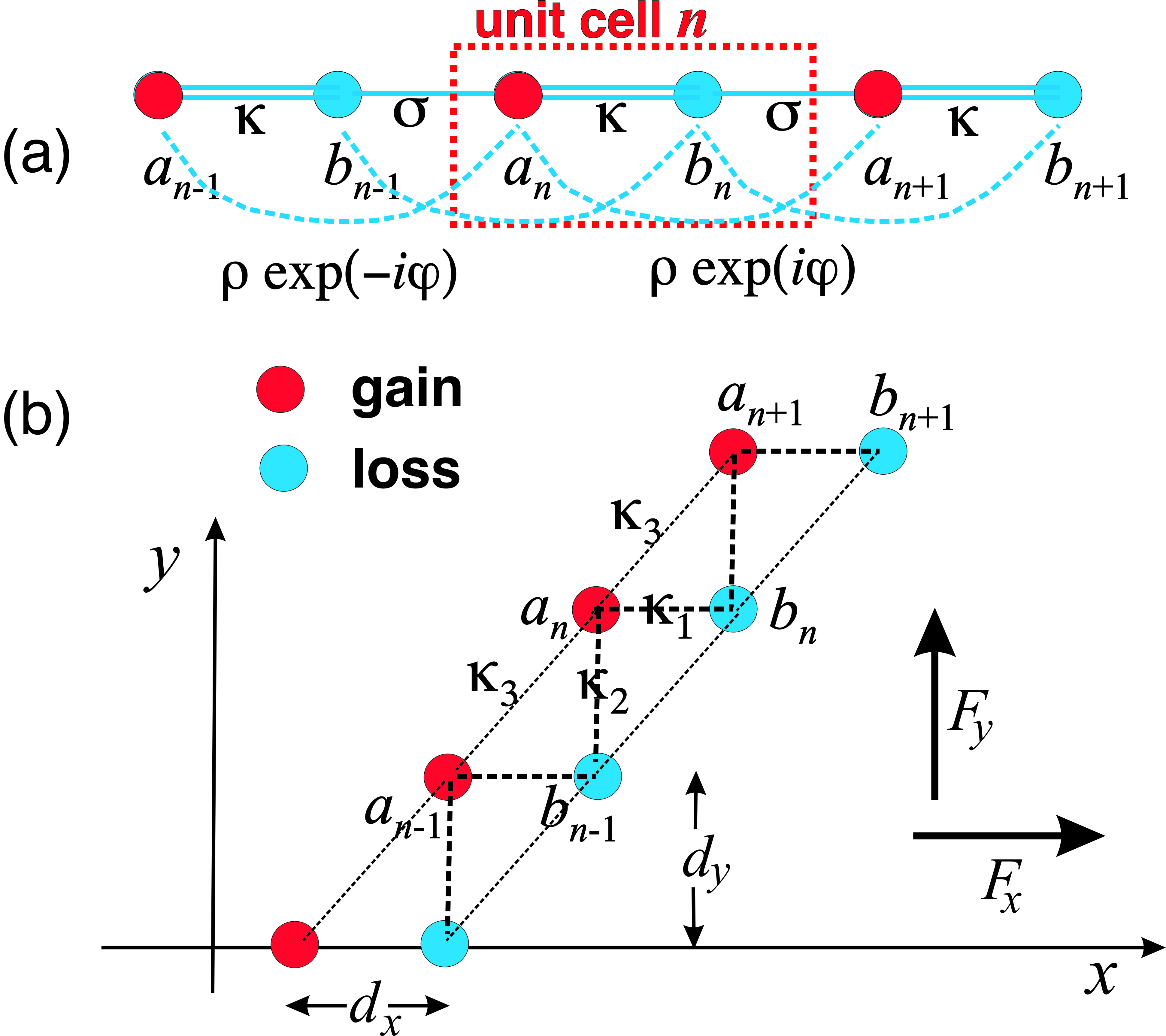}
\caption{(Color online). (a) Schematic of the non-Hermitian extension of the Rice-Mele tight-binding lattice model.  The lattice unit cell contains two sites (a dimer), one with gain and the other with loss.  Next-nearest neighborhood hopping occurs at a rate $\rho \exp(i \varphi)$. Convective transport at the $\mathcal{PT}$ symmetry breaking point is obtained for $\rho \neq 0$ and $\varphi \neq 0,\pi$. (b) Optical realization of the Rice-Mele model in a zig-zag array of optical waveguides with alternating optical gain and loss [cross section in the transverse $(x,y)$ plane]. The optical axis of the array is bent along the paraxial propagation distance $t$. Axis bending realizes an effective combined ac-dc driving of the lattice with forces $F_x(t)$ and $F_y(t)$ along the two transverse directions $x$ and $y$.}
\end{figure}
where $\kappa,  \sigma >0$ are the nearest-neighborhood modulated hopping rates within the unit cell, $\rho \exp(i \varphi)$ is the complex-valued hopping rate of next nearest sites with controlled phase $\varphi$, and $g$ is the  gain/loss rate at alternating sites. The coupled-mode equations (1) and (2) are derived from the tight-binding Hamiltonian 
\begin{eqnarray}
\hat{H} & = & -\sum_n \left( \kappa  \hat{a}_n^{\dag} \hat{b}_{n}+\sigma \hat{a}_n^{\dag} \hat{b}_{n-1}  + H.c. \right) \nonumber \\
& - & \sum_n \left[ \rho \exp(i \varphi) \left( \hat{a}_n^{\dag} \hat{a}_{n+1} +\hat{b}_n^{\dag} \hat{b}_{n+1} \right) + H.c. \right] \;\;\;\;\; \\
& +& ig \sum_n \left( \hat{a}_n^{\dag} \hat{a}_{n}-\hat{b}_n^{\dag} \hat{b}_{n} \right) \nonumber
\end{eqnarray}
which is Hermitian in the limiting case $g=0$ or after replacing $g \rightarrow ig$. The lattice Hamiltonian (3) is invariant under simultaneous parity transformation and time reversal, and can be regarded as a non-Hermitian extension of the Rice-Mele Hamiltonian \cite{Rice,note}, originally introduced to model conjugated diatomic polymers \cite{Rice} and found in other physical systems as well, for example in cold atoms moving in one-dimensional optical superlattices \cite{Bloch}. A possible physical implementation of this Hamiltonian will be discussed in the following subsection. We note that tight-binding lattice models with non-Hermitian terms have been introduced and studied in several recent works \cite{TB1,TB2,TB3,TB4}. In particular, the limiting case $\rho=0$ and $\kappa_1=\kappa_2$ of the Hamiltonian (3) was previously considered in Refs.\cite{Longhi_PRL_09,TB1}, where for the infinitely-extended system  $\mathcal{PT}$ symmetry breaking was shown to occur at $g=g_{th}=0$. As discussed in the next section, this kind of $\mathcal{PT}$ symmetry breaking is always absolute.\\ 
For the general case $\kappa_1 \neq \kappa_2$ and $\rho \neq 0$, the onset of $\mathcal{PT}$ symmetry breaking can be readily determined by analytical calculation of the energy spectrum of the Hamiltonian (3). To this aim, let us search for a solution to Eqs.(1) and (2) in the form of Bloch-Floquet states 
\begin{equation}
\left( 
\begin{array}{c}
a_{n}(t) \\
b_{n}( t)
\end{array}
\right) =  \left( 
\begin{array}{c}
A \\
B 
\end{array}
\right) \exp(-iEt+iqn) \;\;\;
\end{equation}
where $q$ is the quasi-momentum, which is assumed to vary in the interval $(0, 2 \pi)$, and $E=E(q)$ the corresponding energy. 
Substitution of the Ansatz (4) into Eqs.(1) and (2) yields the following homogeneous linear system for the complex amplitudes $A=A(q)$ and $B=B(q)$
\begin{eqnarray}
\left[ E+2 \rho \cos( q+\varphi) -ig \right] A +[\kappa + \sigma \exp(-iq)] B & = & 0 \nonumber \\
\left[ \kappa+\sigma \exp(iq) \right] A+ \left[ E+2 \rho \cos( q+\varphi) +ig \right]  B & = & 0  \;\;\;\;
\end{eqnarray}
which is solvable provided that the determinantal equation 
\begin{equation}
\left|
\begin{array}{cc}
E + 2 \rho \cos (q+\varphi)-ig & \kappa+\sigma \exp(-iq ) \\
 \kappa+\sigma \exp(iq ) & E + 2 \rho \cos (q+\varphi) +ig 
\end{array}
\right|=0 \;\;\;
\end{equation}
is satisfied. This yields the following dispersion relations $E=E_{\pm}(q)$ for the two superlattice minibands
\begin{equation}
E_{\pm}(q)=-2 \rho \cos(q+\varphi) \pm \sqrt{-g^2+\kappa^2+\sigma^2+2 \kappa \sigma \cos q }
\end{equation}
and the following expressions for the amplitudes $A$, $B$ of Bloch-Floquet eigenmodes
\begin{equation}
\left(
\begin{array}{c}
A_{\pm}(q) \\
B_{\pm}(q)
\end{array}
\right)=
\left(
\begin{array}{c}
\kappa+\sigma \exp(-iq) \\
ig-E_{\pm}(q)-2 \rho \cos(q+\varphi)
\end{array}
\right).
\end{equation}
From Eq.(7) it follows that the energy spectrum is entirely real-valued for $g<g_{th}$ with $g_{th} \equiv |\sigma-\kappa|$. In this case, corresponding to the unbroken $\mathcal{PT}$ phase, the energy spectrum comprises two minibands which do not cross. In particular, at $q=\pi$ the two minibands are separated by an energy gap of width $2 \sqrt{g_{th}^2-g^2}$. As $g \rightarrow g_{th}^-$ the gap at $q= \pi$ shrinks and the two minibands touch at $q=\pi$; as $g$ overcomes $g_{th}$, complex-conjugate energies appear near $q=\pi$, which is the signature of $\mathcal{PT}$ symmetry breaking; see Fig.2. It is worth noticing that the group velocity $v_g$ of Bloch modes near $q= \pi$ at the symmetry breaking point, defined by $v_g=(d {\rm {Re}} (E_{\pm})/dq)$, is given by
\begin{equation}
v_g=-2 \rho \sin \varphi
\end{equation}
which does not vanish provided that $\rho \neq 0$, i.e. in the presence of next-nearest neighborhood hopping, and $\varphi \neq 0, \pi$. As it will be shown in Sec.III.B, a non-vanishing and sufficiently large group velocity can cause the $\mathcal{PT}$ symmetry breaking to change from absolute to convective.
\begin{figure*}
\includegraphics[width=14cm]{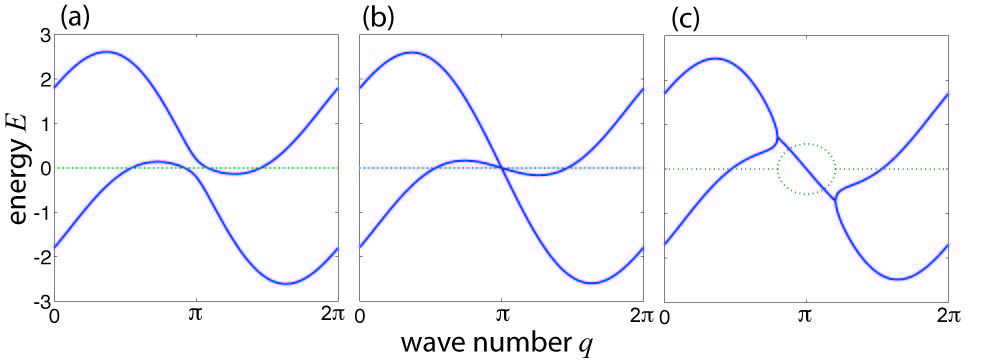}
\caption{(Color online). Energy spectrum of the non-Hermitian Rice-Mele Hamiltonian (3)  (a) in the unbroken $\mathcal{PT}$ phase ($g=0$), (b) at the symmetry breaking point ($g=g_{th}$), and (c) in the broken $\mathcal{PT}$ phase ($g=3g_{th}$). Real and imaginary parts of the energies for the two superlattice minibands  are depicted by the continuous and dashed lines, respectively. Parameter values are $\kappa=1$, $\sigma=0.8$, $\rho=0.6$, and $\varphi= \pi/2$, corresponding to $g_{th}=|\kappa-\sigma|=0.2$.}
\end{figure*}

\subsection{Ac-dc driven lattice model and optical realization of the non-Hermitian Rice-Mele Hamiltonian}
Before discussing the nature of the $\mathcal{PT}$ symmetry breaking for the extended non-Hermitian Rice-Mele Hamiltonian (3), it is worth suggesting possible physical implementations of this model. To realize the Hamiltonian (3), in addition to the non-Hermitian (gain and loss) terms one needs to implement next-nearest neighborhood hoppings with controlled phase $\varphi$. Rather generally, tight-binding lattice models with controlled phase of hopping rates 
can be realized by combined ac-dc forcing. 
Here we briefly propose a photonic realization of the extended Rice-Mele model, based on light transport in a superlattice of evanescently-coupled optical waveguides. The Rice-Mele Hamiltonian (3) can be basically obtained as a limiting case of an ac-driven tight-binding lattice at high modulation frequencies. Another possible physical system where the combined ac-dc driven lattice model could be implemented  is provided by cold atoms trapped in optical superlattices \cite{Bloch}, where gain is introduced via atom injection at alternating sites \cite{kor,BEC}. However, in spite of several theoretical proposals, experimental realizations of $\mathcal{PT}$-symmetric Hamiltonians using ultracold atoms is still missing, and hence we limit here to briefly discuss the photonic system. The optical structure  that we consider is shown in Fig.1(b) and is basically composed by a sequence of evanescently-coupled optical waveguides in a zig-zag geometry with alternating optical amplification (gain) and loss. The waveguides  are displaced in the horizontal ($x$) and vertical ($y$) directions by the distances $d_x$ and $d_y$, respectively.  In the zig-zag geometry, non-negligible evanescent coupling occurs for nearest and next-nearest waveguides \cite{Felix}, with coupling constants (hopping rates) $\kappa_1$, $\kappa_2$ for adjacent guides and $\kappa_3$ for next-nearest guides, as indicated in Fig.1(b).
The values of the coupling constants $\kappa_1$, $\kappa_2$ and $\kappa_3$ are determined by certain overlapping integrals of the optical modes trapped in the waveguides, and they are usually exponentially-decaying functions of waveguide separation. For dielectric waveguides, the coupling constants take real and positive values.
 The difference of couplings $\kappa_1$ and $\kappa_2$ can be controlled by changing the horizontal ($d_x$)  and vertical ($d_y$) distances of waveguides, with $\kappa_1=\kappa_2$ for $d_x \simeq d_y$. For straight waveguides, the array of Fig.1(b) thus realizes the extended Rice-Mele model of Fig.1(a) with $\kappa=\kappa_1$, $\sigma=\kappa_2$, $\rho=\kappa_3$ and $\varphi=0$. 
To realize an effective complex-valued amplitude for the hopping rate between next-nearest neighborhood guides, i.e. $\varphi \neq 0$, we bend the waveguide axis  in both $x$ and $y$ directions along the paraxial propagation distance $t$, so that the optical axis of the array describes a curved path with parametric equations $x=x_0(t)$ and $y=y_0(t)$. Arrays of waveguides with arbitrarily curved axis in three-dimensions can be realized, for example, by the technique of femtosecond laser writing in optical glasses (see, for instance, \cite{Crespi}). In the tight-binding and paraxial approximations, light transport in the superlattice with a bent axis is governed by the following coupled-mode equations (see, for instance, \cite{LonghRev})
\begin{eqnarray}
i \frac{dA_n}{dt} & = & -\kappa_1 B_n -\kappa_2 B_{n-1}-\kappa_3 (A_{n+1}+A_{n-1}) \nonumber \\
 & - & [F_x(t)+F_y(t)]n A_n+igA_n \\
i \frac{dB_n}{dt} & = &  -\kappa_1 A_n -\kappa_2 A_{n+1}-\kappa_3 (B_{n+1}+B_{n-1}) \nonumber \\
 & - & [F_x(t)+F_y(t)]nB_n-F_x(t) B_n-igB_n \;\;\;
\end{eqnarray}
where $A_n$, $B_n$ are the mode amplitudes of light trapped in the alternating waveguides with optical gain and loss, respectively, $g$ is the optical gain/loss coefficient, and 
\begin{equation}
F_x(t)=-\frac{2 \pi n_sd_x}{ \lambda} \frac{d^2x_0}{dt^2} \;, \; \;  F_y(t)=\frac{2 \pi n_sd_y}{ \lambda} \frac{d^2y_0}{dt^2}. \;\;\;
\end{equation}
 account for the axis bending in the horizontal ($x$) and vertical ($y$) directions \cite{LonghRev,LonghiPRL06}. In Eq.(12), $\lambda$ is the wavelength of the propagating light and $n_s$ is the substrate refractive index at wavelength $\lambda$.  Note that Eqs.(10) and (11) describe a  dimerized lattice with external forcing, with $F_x(t)$ and $F_y(t)$ playing the role of the external forces. Note also that, in the absence of axis bending, i.e. for $F_x=F_y=0$, Eqs.(10) and (11)  reproduce the extended Rice-Mele model [Eqs.(1) and (2)] with $\varphi=0$. 
 The equivalence of the driven lattice model [Eqs.(10) and (11)] with the static Rice-Mele lattice model [Eqs.(1) and (2)] with $\varphi \neq 0$ can be established as follows. Let us tailor the axis bending profiles $x_0(t)$ and $y_0(t)$ in the horizonatl and vertical directions to realize the following ac-dc forces $F_x(t)$ and $F_y(t)$ 
 \begin{eqnarray}
 F_x(t) & = & U-(\Gamma \omega) \cos(\omega t + \phi) \nonumber \\
  F_y(t) & = & -U-(\Gamma \omega) \cos(\omega t-\phi),
 \end{eqnarray}
 where $U$,  $\Gamma$ and $\omega$ are real-valued positive parameters. In  our optical waveguide system, the combined ac-dc forcing corresponds to  a sinusoidal axis bending with spatial frequency $\omega$ superimposed to a parabolic path \cite{Dignam}. Note that the sinusoidal bending is not in phase for the horizontal and vertical directions owing to the phase term $\phi$.  
Let us further assume that the following resonance condition 
\begin{equation}
M \omega=U
\end{equation}
is satisfied for some integer $M$, and let us  introduce the amplitudes $a_n$, $b_n$ via the gauge transformation
\begin{eqnarray}
A_n (t)& = & a_n (t) \exp[i \varphi n+i n \Phi(t)] \\
B_n (t) & = & b_n (t) \exp[i \varphi n +i \beta + i n \Phi(t)+i \Theta(t)]
\end{eqnarray}
where we have set
 \begin{equation}
 \Phi (t)=\int_0^t dt' [F_x(t')+F_y(t')] \; , \; \; \Theta(t)=\int_0^t dt' F_x(t'),
 \end{equation}
 $\beta=M \phi-\Gamma \sin \phi$, and
\begin{equation}
\varphi=2 M \phi + M \pi 
\end{equation}
 Substitution of Eqs.(15) and (16) into Eqs.(10,11) yields a system of coupled-equations for the amplitudes $a_n(t)$ and $b_n(t)$ with time-periodic coefficients of period $T=2 \pi / \omega$. As shown in the Appendix A, if the system is observed at discrete times $\tau=0,T,2T,3T,...$, the evolution of the amplitudes $a_n(\tau)$, $b_n(\tau)$ can be mapped into the dynamics of an effective static lattice (i.e. with time-independent hopping rates) which sustains two minibands with dispersion relations $E_{\pm}(q)$ given by the quasi-energies of the  original time-periodic system. In particular,  in the large modulation limit $\omega \gg \kappa_{1}, \kappa_2, \kappa_3,g$,  i.f. for $T \rightarrow 0$, it can be shown (see Appendix A) that the a-dc driven lattice model exactly reproduces the Rice-Mele static model [Eqs.(1) and (2)] with effective hopping rates given by 
\begin{eqnarray}
\kappa & = & \kappa_1J_M(\Gamma) \\
\sigma & = & \kappa_2 J_M(\Gamma) \\
\rho & = & \kappa_3 J_0(2 \Gamma \cos \phi)
\end{eqnarray}
and with the phase $\varphi$ given by Eq.(18), where $J_n$ is the Bessel function of first kind and order $n$. Therefore, the zig-zag waveguide array of Fig.1(b) with alternating optical gain and loss and with a suitable axis bending effectively realizes the extended Rice-Mele lattice model of Fig.1(a)  with a non-vanishing advective term $\varphi$ and with controlled hopping rates $\kappa$, $\sigma$, $\rho$. 

\section{Convective and absolute $\mathcal{PT}$ symmetry breaking}
In this section we introduce the notion of convective and absolute $\mathcal{PT}$ symmetry breaking, inspired by the concepts of convective and absolute unstable flows in hydrodynamics \cite{rev1,flow1,flow2}, and then we apply such concepts to the non-Hermtiian Rice-Mele and ac-dc driven  models presented in Sec.II.
\subsection{Definition of absolute and convective  $\mathcal{PT}$ symmetry breaking for a periodic potential}
In this subsection we present the rather general definition of convective and absolute $\mathcal{PT}$ symmetry breaking for a continuous system in one spatial dimension $x$, described by a $\mathcal{PT}$-invariant Hamiltonian $\hat{H}=-\partial^2_x+V(x)$ with a potential $V(x)=V_R(x)+i g V_I(x)$, where $V_R(-x)=V_R(x)$ and $V_I(-x)=-V_I(x)$ are the real and imaginary parts of the potential and $g \geq 0$ is a real-valued parameter that measures the strength of the non-Hermitian part of the potential.  The concept of convective and absolute $\mathcal{PT}$ symmetry breaking is meaningful in case where at the symmetry breaking point complex-conjugate energies emanate from the continuous spectrum of $\hat{H}$, i.e. the corresponding eigenstates are not normalizable. In fact, if the symmetry breaking arises because of the appearance of pairs of normalizable states with  complex-conjugate energies, the $\mathcal{PT}$ symmetry breaking is always absolute and can not be convective, according to the hydrodynamic definitions of absolute and convective unstable flows briefly mentioned in the introduction section and formally defined below.  An important case where  $\mathcal{PT}$ symmetry breaking arises because of the emergence of extended (non-normalizable) states with complex conjugate energies is the one of a periodic potential, $V(x+d)=V(x)$. In this case, the energy spectrum is absolutely continuous and composed by energy bands. We assume that the energy spectrum of $\hat{H}$ is entirely  real-valued for $g \leq g_{th}$, corresponding to the unbroken $\mathcal{PT}$ phase, whereas complex-conjugate energies appear for $g>g_{th}$, where $g_{th} \geq 0$ determines the symmetry breaking point. For example, for the potential $V_R(x)=\cos(2 \pi x/d)$ and $V_I(x)=\sin (2 \pi x /d)$ $\mathcal{PT}$ symmetry breaking is attained at $g_{th}=1$ \cite{Makris_PRL_08,LonghiPRA10,Longhi_JPA_2011,Graefe11}.  Let us then consider an initial wave packet  $\psi(x,0)$ at time $t=0$, and let $\psi(x,t)=\exp(-i \hat{H}t) \psi(x,0)$ be the evolved wave packet at successive time $t$. In the unbroken $\mathcal{PT}$ phase, one has $\psi(x,t) \rightarrow 0$ as $t \rightarrow \infty$ at any fixed position $x$ owing to delocalization of the wave packet in the lattice. However, in the broken $\mathcal{PT}$ phase, i.e. for $g>g_{th}$, owing to the appearance of complex energies the wave packet $\psi(x,t)$ is expected to secularly grow as $t \rightarrow \infty$. According to the definitions of unstable flows in hydrodynamic systems \cite{rev1,flow1}, the $\mathcal{PT}$ symmetry breaking is said to be {\it absolute} if $\psi(x,t) \rightarrow \infty$ at $x=0$ (or at any fixed position $x=x_0$), whereas it is said to be convective if  $\psi(x,t) \rightarrow \infty$ along the ray $x=vt$ for some drift velocity $v$, but $\psi(x,t) \rightarrow 0$ at $x=0$ (or at any fixed position $x=x_0$). The physics behind the definition of absolute and convective unstable flows is rather simple and is visualized in Fig.3. In the convectively unstable regime, an initial wave packet (perturbation) drifts in the laboratory reference frame with some velocity $v$, and along the ray $x=vt$, i.e. in the reference frame moving with the wave packet, the perturbation secularly grows with time. The drift velocity $v$ is basically determined by the wave packet group velocity at the quasi-momentum $k=k_s$ where the maximum growth rate (i.e. largest imaginary part of the energy) occurs. However, at a fixed position $x=x_0$ (e.g. $x_0=0$), the perturbation $\psi(x_0,t)$ can grow only transiently, but finally it vanishes as $t \rightarrow \infty$  owing to the (possibly fast) drift of the growing wave packet [see Fig.3(a)]. Conversely, in the absolutely unstable regime the perturbation grows so fast that, even in the presence of an advective term (a drift),   at a fixed spatial position $x_0$ the perturbation $\psi(x_0,t)$ grows indefinitely with time [see Fig.3(b)]. To determine whether the $\mathcal{PT}$ symmetry breaking is convective or absolute, let us consider the Hamiltonian $\hat{H}$ with $g>g_{th}$, and let us consider an initial wave packet given by a superposition of  Bloch-Floquet modes $\phi_k(x)=u_k(x) \exp(ikx)$ with energy $E=E(k)$, i.e. $\hat{H} \phi_k(x)=E(k) \phi_k(x)$, with $u_k(x+d)=u_k(x)$ and which the quasi-momentum $k$ that varies from $-\infty$ to $\infty$ to account for all the lattice bands (extended band representation). The wave packet then evolves according to the relation 
\begin{equation}
\psi(x,t)= \int_{-\infty}^{\infty} dk F(k) u_k(x) \exp[ikx-iE(k)t]
\end{equation}
where $F(k)$ is the spectrum of excited Bloch-Floquet modes.
Along the ray $x=vt$ one has
\begin{equation}
\psi(t)= \int_{-\infty}^{\infty} dk F(k) u_k(vt) \exp[ikvt-iE(k)t].
\end{equation}
The determination of the nature (absolute or convective) of the $\mathcal{PT}$ symmetry breaking  entails the estimation of the asymptotic behavior of $\psi(t)$ as $ t \rightarrow \infty$.   Since  $u_k(x)$ is a limited and periodic function of $x$, we can study the asymptotic behavior of the associated wave packet 
\begin{equation}
\psi_1(t)= \int_{-\infty}^{\infty} dk F(k) \exp[ikvt-iE(k)t]
\end{equation}
obtained by dropping the term $u_k(vt)$ under the integral in Eq.(23). In fact, it can be readily shown that ${\rm  lim \; sup}_{t \rightarrow \infty } |\psi (t)| \rightarrow \infty$  ($ \rightarrow 0$) if and only if ${\rm  lim \; sup}_{t \rightarrow \infty } |\psi_1(t)| \rightarrow \infty$ ($ \rightarrow 0$). Note that for the determination of the asymptotic behavior of $\psi_1(t)$ we only need to evaluate the integral on the right hand side of Eq.(24) for those values
of $k$ for which ${\rm Im} \{ E(k) \} \geq 0$, the other modes giving
no contribution  (they are surely decaying).The asymptotic behavior of $\psi_1(t)$ as $t \rightarrow \infty$ can be determined, under certain conditions which are generally satisfied,  by the saddle-point (or steepest descend) method \cite{rev1}. This entails analytic continuation of the function $E(k)$ is the complex $k$ plane and, using the Cauchy theorem, the deformation of the path of the integral  along a suitable contour which crosses a (dominant) saddle point $k_s$ of $E(k)-kv$ in the complex
plane, along the direction of the steepest descent \cite{rev1,flow1,flow2}. The
asymptotic behavior of the integral is then given by the value
of the exponential part of the integrand calculated at the
saddle point. More precisely, for a saddle point of order $n \geq 2$, i.e. for which $E(k)=E(k_s)+v(k-k_s)+(d^n E/dk^n)_{k_s}(k-k_s)^n+o((k-k_s)^n)$,
for $t \rightarrow \infty$ one has \cite{steep}
\begin{eqnarray}
\psi_1(t) & \sim & \frac{F(k_s)}{|t (d^n E/dk^n)_{k_s}|^{1/n}} (n!)^{1/n} \Gamma \left( \frac{1}{n} \right) \nonumber \\
& \times & \exp  [it vk_s \pm i  \pi /(2n) ] \exp[-it E(k_s)]
\end{eqnarray}  
where the saddle point $k_s$ in the complex plane is determined from the equation
\begin{equation}
\left( \frac{dE}{dk} \right)_{k_s}=v.
\end{equation}
The decay or secular growth of $\psi_1(t)$ thus depends on the sign of the imaginary part of $E(k)$ at the saddle point $k=k_s$. It can be readily shown that, for $g>g_{th}$, there is always a velocity $v=v_s$ for which the solution $k_s$ to Eq.(26) is real-valued and corresponds to the maximum growth rate  [i.e. the maximum of ${\rm Im} (E(k))>0$], so that along the ray $x=v_s t$ the amplitude $\psi_1(t)$ shows a secular growth. To determine whether the symmetry breaking is either convective or absolute, we should consider the asymptotic behavior of $\psi_1(t)$ for $v=0$, which is determined by the sign of the imaginary part of $E(k)$ at the saddle point $k=k_s$ obtained from Eq.(26) with $v=0$. Hence,  the $\mathcal{PT}$ symmetry breaking is absolute if ${\rm Im} \{ E(k_s) \}>0$, whereas it is convective if  ${\rm Im} \{ E(k_s) \} \leq  0$, where the saddle point $k_s$ is determined from the equation $(dE/dk)_{k_s}=0$. As a general rule of thumb,  for $g$ larger but close the $\mathcal{PT}$ symmetry breaking threshold, indicating by $k_s$ the quasi momentum on the real axis with maximum growth rate, i.e. that maximizes ${\rm Im}(E(k))$ for $k$ real, the $\mathcal{PT}$ symmetry breaking is absolute if the group velocity $v_g$ at $k=k_s$, given by $v_s=(d {\rm Re}(E) /dk)_{k_s}$, vanishes, whereas is it expected to be convective for a nonvanishing (and possibly large) value of $v_s$. Physically, the latter regime corresponds to the case where, owing to a non-vanishning group velocity, the unstable growing Bloch-Floquet mode is  advected away, for an observer at rest,  fast enough that it decays in time when observed at a fixed spatial position.
   
\begin{figure}
\includegraphics[width=8.3cm]{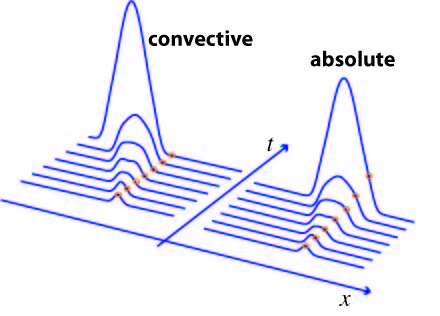}
\caption{(Color online). Schematic of wave packet evolution in the convective and absolute  $\mathcal{PT}$ symmetry breaking regimes. The dotted lines show the evolution of the wave packet along the path $x=0$.}
\end{figure}

\begin{figure}
\includegraphics[width=8.3cm]{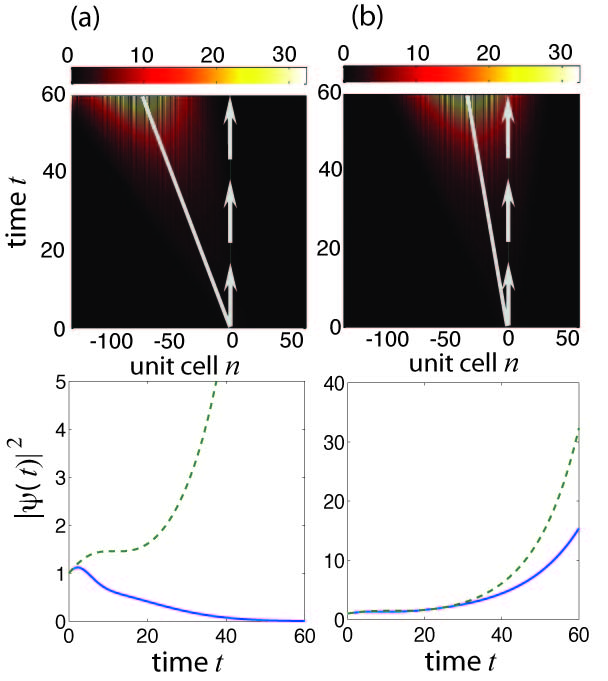}
\caption{(Color online). Numerically-computed wave packet evolution (snapshots of $|\psi(n,t)|^2$)  for the Rice-Mele Hamiltonian Eq.(3) in (a) convective, and (b) absolute $\mathcal{PT}$ symmetry breaking regimes. The lower panels show the detailed temporal evolution of the occupation probabilities of the lattice sites $a_n$ along the two rays indicated in the upper panels by the tilted solid curve (path $n=v_gt$ with maximum growth, dashed curve) and by the vertical arrows (path $n=0$, solid curves). Parameter values are given in the text.}
\end{figure}

\subsection{Absolute and convective  $\mathcal{PT}$ symmetry breaking for the non-Hermitian Rice-Mele Hamiltonian}
In this subsection we describe in details the nature of the $\mathcal{PT}$ symmetry breaking for the extended Rice-Mele Hamiltonian defined by Eq.(3). As shown in Sec.II.A, the superlattice comprises two minibands, with dispersion relations $E_{\pm}(q)$ and corresponding Bloch-Floquet modes defined by Eqs.(7) and (4,8), respectively. After setting $\psi(n,t)=(a_n(t),b_n(t))^T$, let us consider the propagation of an initial wave packet $\psi(n,0)$ in the lattice, which is assumed to be given by a superposition of Bloch-Floquet modes belonging to the two minibands  with  spectral functions $F_{\pm}(q)$. The evolved wave packet at time $t$ is then given by
\begin{eqnarray}
\psi(n,t) & = & \int_{0}^{2 \pi} dq F_+(q) \phi_+(q) \exp[iqn-iE_{+}(q)t] \nonumber \\
& + & \int_{0}^{2 \pi} dq F_-(q) \phi_-(q) \exp[iqn-iE_{-}(q)t] \;\;\;\;\;\;
\end{eqnarray}
where we have set $\phi_{\pm}(q)=(A_{\pm}(q),B_{\pm}(q))^T$. As shown in Sec.II.A, $\mathcal{PT}$ symmetry breaking occurs when the gain/loss parameter $g$ is increased to overcome the threshold value $g_{th}=|\kappa-\sigma|$. Correspondingly, complex conjugate energies appear for a wave number $q$ close to $q_0=\pi$ [see Fig.2(c)].   Note that, since ${\rm Im} \{ E(q) \} \geq 0$ for one miniband and  ${\rm Im} \{ E(q) \} \leq 0$ for the other miniband, one of the two integrals on the right hand side of Eq.(27) decays toward zero as $t \rightarrow \infty$, and therefore we can limit to consider the contribution arising from the other integral involving unstable modes. Assuming, for the sake of definiteness, ${\rm Im} \{ E_+(q) \} \geq 0$ and ${\rm Im} \{ E_-(q) \} \leq 0$, one has
\begin{equation}
\psi(n,t) \sim  \int_{0}^{2 \pi} dq F_+(q) \phi_+(q) \exp[iqn-iE_{+}(q)t]
\end{equation}
as $t \rightarrow \infty$. The asymptotic form of the integral on the right hand side of Eq.(28) along the ray $n=vt$ can be estimated by the saddle point method and takes a form similar to the one given by Eq.(25). According to the analysis presented in Sec.III.A, the $\mathcal{PT}$ symmetry breaking is thus convective if ${\rm{Im}} \{ E_+(q_s)\} \leq  0$ , whereas it is absolute 
for  ${\rm{Im}} \{ E_+(q_s)\}>0$, where $q_s$ is the dominant saddle point obtained from the equation $(dE_+/dq)_{q_s}=0$, i.e. 
\begin{equation}
\frac{2 \rho}{\kappa \sigma} \left (\cos \varphi \sin q_s \sin \varphi \cos q_s \right)=\frac{\sin q_s}{-\epsilon^2+2 \kappa \sigma(1+ \cos q_s)}.
\end{equation}
In Eq.(29) we have set $\epsilon^2=g^2-g_{th}^2$, which provides a measure of the distance from the $\mathcal{PT}$ symmetry breaking point. To simplify our analysis, let us consider the case where the gain/loss parameter $g$ is larger but close to its threshold value $g_{th}$, so that $\epsilon^2$ is a small quantity. In this case the solutions to Eq.(29)  can be determined analytically by an asymptotic analysis in the small parameter $\epsilon$. The calculations are detailed in the Appendix B. The main result of the calculations is that the $\mathcal{PT}$ symmetry breaking is {\it convective} for 
\begin{equation}
|v_g|> \sqrt{\sigma \kappa}
\end{equation}
\begin{figure}
\includegraphics[width=8.3cm]{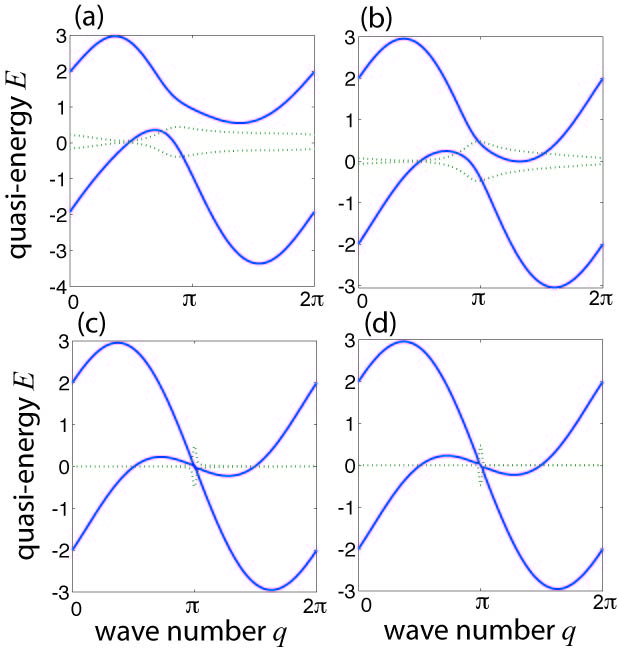}
\caption{(Color online). Numerically-computed quasi-energy minibands $E_{\pm}(q)$ for the ac-dc driven lattice model [Eqs.(10) and (11)] for increasing values of the modulation frequency $\omega$: (a) $\omega=6$,  (b) $\omega=15$, and  (c) $\omega=150$. The other parameter values are given in the text. In (d) the energy minibands of the static Rice-Mele lattice are shown, that correspond to the asymptotic limit $\omega \rightarrow \infty$. Solid curves refer to the real part of $ E_{\pm}(q)$, whereas the thin dotted curves to the imaginary part of $ E_{\pm}(q) $. For the sake of clearness,  the imaginary part of $ E_{\pm}(q) $ has been multiplied by a factor of 10.}
\end{figure}
\begin{figure}
\includegraphics[width=8.3cm]{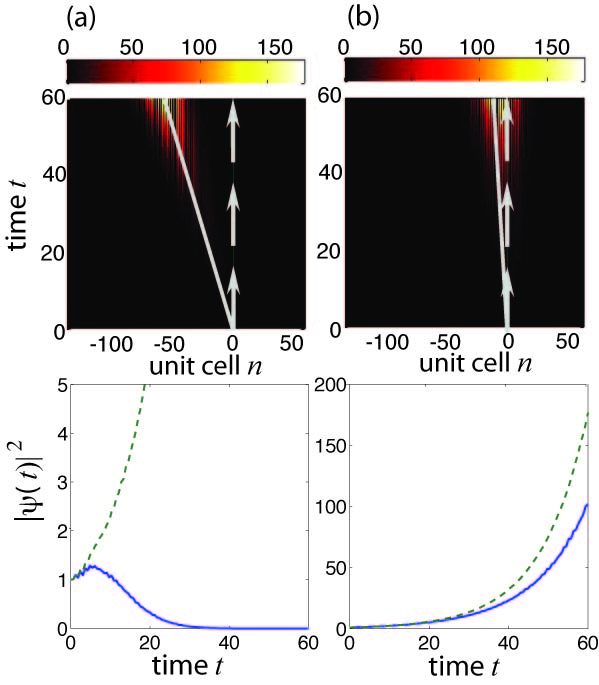}
\caption{(Color online). Numerically-computed wave packet evolution (snapshots of $|\psi_n(t)|^2$)  for the ac-dc-driven lattice model in (a) convective, and (b) absolute $\mathcal{PT}$ symmetry breaking regimes. The lower panels show the detailed temporal evolution of the occupation probabilities of the lattice sites $a_n$ along the two paths indicated in the upper panels by the tilted solid curve (path $n=v_gt$ with maximum growth, dashed curve) and by the vertical arrows (path $n=0$, solid curves). Parameter values are given in the text.}
\end{figure}
 whereas it is {\it absolute} in the opposite case $|v_g| \leq  \sqrt{\sigma \kappa}$, where $v_g=-2 \rho \sin \varphi$ is the group velocity  at the symmetry breaking point  of the most unstable mode with wave number $q=\pi$ [see Eq.(9)]. Hence, as expected, a sufficiently large advective term in the extended Rice-Mele Hamiltonian can change the $\mathcal{PT}$ symmetry breaking from absolute to convective. Note that for $\rho=0$ or $\rho \neq$ but real-valued, the symmetry breaking is always absolute. As discussed in the next subsection, an important physical implication of the convective (rather than absolute) $\mathcal{PT}$ symmetry breaking is that the unbroken $\mathcal{PT}$ phase can be restored in the convectively regime by making the region of alternating gain and loss sites in the lattice {\it spatially limited} rather than extended. 

\subsection{Numerical results}
We checked the predictions of the theoretical analysis and the transition form absolute to convective $\mathcal{PT}$ symmetry breaking induced by advection for both the static Rice-Mele lattice of Fig.1(a) and the ac-dc driven lattice of Fig.1(b) by direct numerical simulations. 
As an example, in Fig.4 we depict the evolution of a wave packet in the Rice-Mele lattice with advective term ($\rho \neq 0$, $\varphi \neq 0, \pi$), showing the transition from convective [Fig.4(a)]  to absolute [Fig.4(b)] $\mathcal{PT}$ symmetry breaking. The numerical results are obtained by solving the coupled-mode equations (1) and (2) using an accurate fourth-order variable-step Runge-Kutta method assuming as an initial condition a Gaussian wave packet with carrier wave number $q_0=\pi$ at lattice sites $a_n$ solely, namely $a_n(0)=\exp[-2(n/w)^2+iq_0n] $ and $b_n(0)=0$, where $w$ is the size of the wave packet. Such an initial condition mainly excites (unstable) Bloch-Floquet modes with imaginary energy at wave numbers $q$ close to the most critical one $q=q_0=\pi$. Parameter values used in the simulations are $\kappa=\sigma=1$ (corresponding to $g_{th}=0$), $\varphi=\pi/2$, $g=0.05$ and $\rho=0.7$ in Fig.4(a), and $\rho=0.3$ in Fig.4(b). In Fig.4(a), the condition $|v_g|>\sqrt{\sigma \kappa}$ is satisfied and, according to the analysis of Sec.III.B, the symmetry breaking is of convective nature. In fact, while the wave packet $|\psi (t)|^2$ secularly grows along the ray $n=v_g t$, it decays when observed at a fixed spatial position (e.g. $n=0$), as shown in the lower panel of Fig.4(a). Conversely, in Fig.4(b) the advective term in the Rice-Mele Hamiltonian is lowered so that $|v_g|$ is smaller than $\sqrt{\sigma \kappa}$: in this case the symmetry breaking is absolute, as clearly shown in the lower panel of Fig.4(b).\\
A similar transition from absolute to convective $\mathcal{PT}$ symmetry-breaking for increasing advection is observed in the ac-dc driven lattice model of Fig.1(b) presented in Sec.II.B. As shown in the Appendix A, the dynamical properties of the ac-dc driven lattice at discretized times $\tau=0,T,2T,...$ can be mapped into the ones of a static lattice with an energy band structure that is determined by the quasi-energy spectrum $E(q)$ of the ac-dc driven lattice.  In particular, at large modulation frequencies the driven lattice model, defined by Eqs.(A1) and (A2), exactly reproduces the Rice-Mele model with effective hopping rates $\kappa$, $\sigma$, $\rho$ and phase $\varphi$ given by Eqs.(18-21). As an example, in Figs.5(a-c) we show the numerically-computed quasi energies of the two minibands (real and imaginary parts) for  the ac-dc driven lattice above the $\mathcal{PT}$ symmetry breaking point for parameter values $\kappa_1=\kappa_2=2.1124$, $\kappa_3=1.4784$, $M=1$, $\Gamma=1.109$, $\phi=-\pi/4$, $g=0.05$ and for increasing values of the modulation frequency $\omega$. Parameter values have been chosen such as to reproduce, at large modulations frequencies, the static Rice-Mele lattice with parameters as in Fig.4(b). The quasi-energies have been obtained by numerical computation of the Floquet exponents for the eigenvalue problem defined by Eqs.(A6) and (A7) given in the Appendix. For comparison, in Fig.5(d) the minibands of the   static Rice-Mele lattice with parameters of Fig.4(b) are also depicted. According to the theoretical analysis, in the high modulation regime the quasi-energy spectrum of the driven lattice asymptotically reproduces the spectrum of the static Rice-Mele model [compare Fig.5(c) and (d)]. At low or moderate values of the modulation frequency $\omega$, deviations from the two models can be clearly appreciated [compare Figs.5(a) and (b) with Fig.5(d)]. In particular, the driven lattice model at low modulation frequencies shows a wider range of wave numbers with complex energies, and the real part of the quasi energies for the two minibands are not degenerate. Nevertheless, the transition from convective to absolute symmetry breaking, which is basically related to the value of the group velocity (the derivative of the real part of the quasi-energy) of the unstable mode at the symmetry breaking point, can be observed even at moderate modulation frequencies. This is shown, as an example, in Fig.6, where we depict the numerically-computed evolution of the same initial Gaussian wave packet  as in Fig.4 but in the ac-dc driven lattice for a modulation frequency $\omega=15$ and for $\kappa_3=1.4784$ [Fig.6(a)], corresponding to a convective symmetry breaking, and $\kappa_3=0.6336$ [Fig.6(b)], corresponding to absolute $\mathcal{PT}$ symmetry breaking.\par 
\begin{figure}
\includegraphics[width=8.3cm]{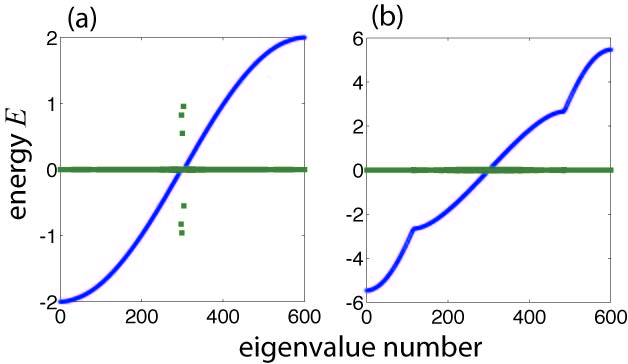}
\caption{(Color online). Numerically-computed energy spectrum (real and imaginary parts) of the Rice-Mele Hamiltonian (32) with a finite number of unit cells with gain and loss regions for (a) $\rho=0$, and (b) $\rho=2$, $\varphi=\pi/2$. The other parameter values are: $\kappa=\sigma=1$ and $g=0.5$. For the sake of clearness the imaginary part of the energies (square points) are multiplied by a factor of 2. The total number of unit cells of the lattice is $N+1=301$, and the eigenvalues are ordered for increasing values of the real part of the energy. The number of dimers with loss and gain is $N_g+1=21$, and they are located at the center of the lattice.}
\end{figure}
As a final comment, it is worth discussing a physically relevant implication of convective versus absolute $\mathcal{PT}$ symmetry breaking.  In the convectively unstable regime, the growing wave packet drifts in the laboratory reference frame fast enough that locally (i.e. at a fixed spatial position) it is observed to decay in spite of  its growth in a moving reference frame (see Figs.3 and 4). Let us now consider a lattice with a spatially confined (rather than infinitely extended) region of unit cells with gain and loss. In the convective regime, advection pushes the wave packet far from the "non-Hermitian" region of the lattice,  and hence after a transient the wave packet ceases to grow. Conversely, in the absolute symmetry breaking regime it is expected to grow indefinitely even for a spatially-finite extension of unit cells with gain and loss. Such a simple physical picture suggests that in the convectively unstable regime $\mathcal{PT}$ symmetry (i.e. an entirely real-valued energy spectrum) may be restored when the gain/loss region in the lattice is spatially limited. We checked such a prediction by considering the Rice-Mele lattice Hamiltonian (3) with a spatially-dependent gain/loss term vanishing at infinity, namely
\begin{eqnarray}
\hat{H} & = & -\sum_{n=-\infty}^{\infty} \left( \kappa  \hat{a}_n^{\dag} \hat{b}_{n}+\sigma \hat{a}_n^{\dag} \hat{b}_{n-1}  + H.c. \right) \nonumber \\
& - & \sum_n \left[ \rho \exp(i \varphi) \left( \hat{a}_n^{\dag} \hat{a}_{n+1} +\hat{b}_n^{\dag} \hat{b}_{n+1} \right) + H.c. \right] \;\;\;\;\; \\
& +& i \sum_{n=-\infty}^{\infty}g_n  \left( \hat{a}_n^{\dag} \hat{a}_{n}-\hat{b}_n^{\dag} \hat{b}_{n} \right) \nonumber
\end{eqnarray}
where $g_n \rightarrow 0$ as $n \rightarrow \infty$. 
In particular, we numerically computed the energy spectrum of $\hat{H}$ by considering a square-wave profile of $g_n$, i.e. $g_n=g$ for $|n| \leq N_g/2$ and $g_n=0$ otherwise. This case corresponds to a central lattice section comprising $(N_g+1)$ dimers with gain and loss (i.e. locally non-Hermitian), and and abrupt transition to two outer lattice sections with locally-Hermitian dimers (i.e. $g_n=0$). In the numerical simulations, the total number of unit cells $(N+1)$ is finite, corresponding to truncation of the outer lattice sections. As an example, in Fig.7 we show the numerically-computed energies of the truncated lattice described by the Hamiltonian 
\begin{eqnarray}
\hat{H} & = & -\sum_{n=-N/2}^{N/2} \left( \kappa  \hat{a}_n^{\dag} \hat{b}_{n}+\sigma \hat{a}_n^{\dag} \hat{b}_{n-1}  + H.c. \right) \nonumber \\
& - & \sum_{n=-N/2}^{N/2} \left[ \rho \exp(i \varphi) \left( \hat{a}_n^{\dag} \hat{a}_{n+1} +\hat{b}_n^{\dag} \hat{b}_{n+1} \right) + H.c. \right] \;\;\;\;\; \\
& +& ig \sum_{n=-N_g/2}^{N_g/2} \left( \hat{a}_n^{\dag} \hat{a}_{n}-\hat{b}_n^{\dag} \hat{b}_{n} \right) \nonumber
\end{eqnarray}
for $N_g=10$, $N=300$ and for parameter values corresponding to absolute [Fig.7(a)] and convective [Fig.7(b)] $\mathcal{PT}$ symmetry breaking in the extended (i.e. $N,N_g \rightarrow \infty$) limit. Note that, within numerical accuracy, the energy spectrum is entirely real-valued in the convective regime [Fig.7(b)], whereas pairs of complex-conjugate energies persist in the absolute regime [Fig.7(a)]. It should be noted, however, that restoring of the $\mathcal{PT}$ symmetry in the convective regime is not a strict rule, since the interfaces from the outer lattice regions to the inner (non-Hermitian) lattice section might sustain localized (interface) modes with imaginary energies, which can not be predicted by our simple picture. Moreover, it is expected that restoring of the $\mathcal{PT}$ symmetry depends on the choice of the profile $g_n$; for example a  smooth (rather than sharp) transition from the inner (locally non-Hermitian) to the outer (locally  Hermitian)  regions is expected to avoid the appearance of interface states. Symmetry breaking in case of inhomogeneous gain/loss parameter $g_n$  would require a further study, however this goes beyond the scope of the present work.

\section{Conclusions}
In this work  we have introduced the concepts of convective and absolute $\mathcal{PT}$ symmetry breaking for wave transport in periodic complex potentials, inspired by the hydrodynamic concepts of convective and absolute instabilities in open flows. In particular, we have investigated analytically and numerically the transition from absolute to convective   $\mathcal{PT}$ symmetry breaking in two tight-binding lattice models: a non-Hermitian extension of the Rice-Mele dimerized lattice, originally introduced to model conjugated diatomic polymers, and an ac-dc driven lattice, which reproduces the Rice-Mele model in the large modulation frequency limit. In the context of spatially-extended dissipative dynamical systems,
$\mathcal{PT}$  symmetry breaking can be viewed as a phase transition from a marginally stable state (the unbroken $\mathcal{PT}$ phase) to an unstable state (the broken $\mathcal{PT}$ phase). The instability arises because of the appearance of pairs of complex-conjugate energies in the broken $\mathcal{PT}$ phase.  The distinction between convective and absolute $\mathcal{PT}$ symmetry breaking arises when considering the evolution of a wave packet in the broken $\mathcal{PT}$ phase: while in the absolute symmetry breaking case the wave packet amplitude observed at a fixed spatial position secularly grows in time,   in the convective symmetry breaking case the amplitude grows in a reference frame moving at some drift velocity, however it decays when observed at a fixed spatial position, i.e. for an observer at rest. A convective regime is generally found when the unstable modes have a group (drift) velocity large enough that at a fixed spatial position the wave packet decay due to the drift   overcomes the growth due to the instability. The nature (either absolute or convective) of the $\mathcal{PT}$ symmetry breaking is basically determined by the sign of the imaginary part of the energy (for static lattices) or quasi-energy (for periodically-driven lattices) at the dominant band saddle point in complex plane. An interesting application of the concepts of convective and absolute symmetry breaking is found when considering a spatially-limited region of gain/loss in the system, i.e when the periodicity of the system is broken and the imaginary part of the potential is confined to a limited region of space. Owing to the fast drift of a wave packet in the convective regime, after a transient the wave packet escapes from the imaginary potential region and thus it ceases to grow. This means that the instability is only transient, i.e. we expect that $\mathcal{PT}$ symmetry is restored in the convective regime when the imaginary potential is spatially confined. This is not the case of the absolute symmetry breaking regime, where the broken $\mathcal{PT}$ phase is expected to persist even for a spatially-limited imaginary potential. Other possible applications and developments of the hydrodynamic concepts of convective and absolute instabilities can be foreseen into the rapidly growing field of wave transport in $\mathcal{PT}$-symmetric quantum and classical systems. For example, like for hydrodynamic and dissipative optical systems \cite{rev1,Santa}, interesting effects (like the appearance of noise-sustained structures \cite{Santa}) might be envisaged for convective  $\mathcal{PT}$ symmetry breaking in presence of classical or quantum noise \cite{uffa}.

\appendix
\section{Floquet analysis of the ac-dc driven lattice and effective static lattice model}
In this Appendix we present a Floquet analysis of the driven lattice model defined by Eqs.(10) and (11) with time-periodic coefficients and show that, at discretized times, it behaves like an effective static lattice with a band structure that is determined by the quasi-energy spectrum of the driven lattice. To this aim, let us note that, after the gauge transformation defined by Eqs.(15) and (16) given in the text, the evolution of the amplitudes $a_n(t)$, $b_n(t)$ is governed by the following linear system of equations
\begin{eqnarray}
i \frac{da_n}{dt} & = & -\kappa_1 F(t) b_{n}-\kappa_2 G(t) b_{n-1}-\kappa_3H(t) a_{n+1} \nonumber \\
& - & \kappa_3 H^*(t) a_{n-1}+iga_n \\
i \frac{db_n}{dt} & = & -\kappa_1 F^*(t) a_{n}-\kappa_2 G^*(t) a_{n+1}-\kappa_3H(t) b_{n+1} \nonumber \\
& - & \kappa_3 H^*(t) b_{n-1}-igb_n
\end{eqnarray}
with time-dependent coefficients $F(t)$, $G(t)$ and $H(t)$ given by
\begin{eqnarray}
F(t) & = & \exp \left[ i \beta +i \Theta (t) \right] \nonumber \\
G(t) & = & \exp \left[ i \beta -i \varphi +i \Theta(t)-i \Phi(t) \right] \\
H(t) & = & \exp \left[ i \varphi + i \Phi(t) \right]. \nonumber
\end{eqnarray}
In the previous equations, the functions $\Theta(t)$ and $\Phi(t)$ and constant parameters $\varphi$ and $\beta$ are defined by Eqs.(17) and (18)  given in the text. For the driving terms $F_x$, $F_y$ defined by Eq.(13), one has explicitly
\begin{eqnarray}
F(t) & = & \exp \left[ iM \phi +i M \omega t -i \Gamma \sin(\omega t + \phi)  \right] \nonumber \\
G(t) & = & \exp \left[ -i M (\phi+ \pi) +iM \omega t +i \Gamma \sin (\omega t - \phi)  \right]  \; \; \; \; \; \; \; \\
H(t) & = & \exp \left[ i M ( 2 \phi+\pi) -2 i \Gamma \cos \phi \sin (\omega t) \right]. \nonumber
\end{eqnarray}
where we assumed the resonance condition $U= M \omega$.  Since the coefficients $F(t)$, $G(t)$ and $H(t)$ are periodic in time with period $T=2 \pi / \omega$, the solution to Eqs.(A1) and (A2)  can be obtained from Floquet theory of liner periodic systems. Specifically, the  general solution to Eqs.(A1) and (A2) is given by an arbitrary superposition of Bloch-Floquet states
\begin{equation}
\left( 
\begin{array}{c}
a_n(q,t) \\
b_n(q,t)
\end{array}
\right)=
\left( 
\begin{array}{c}
A(q,t) \\
B(q,t)
\end{array}
\right) \exp \left[ iqn-i E(q) t \right]
\end{equation}
where  $q$ is the wave number (quasi-momentum), which varies in the range $(0, 2 \pi)$, $E(q)$ is the quasi-energy, with $-\omega/2 \leq {\rm {E}}(q)< \omega/2$, and $A(q,t)$, $B(q,t)$ are periodic in time with period $T$. The quasi-energy $E(q)$ and corresponding Floquet states $(A(q,t),B(q,t))^T$ are found by solving the eigenvalue problem
\begin{eqnarray}
E(q) A & = & -i \frac{dA}{dt}-\kappa_3[H \exp(iq)+H^* \exp(-iq)]A \nonumber \\
& + & igA-[\kappa_1 F + \kappa_2 G \exp(-iq)]B \\
E (q) B & = & -i \frac{dB}{dt}-\kappa_3[H \exp(iq)+H^* \exp(-iq)]B \nonumber \\
& - & igB-[\kappa_1 F^* + \kappa_2 G^* \exp(iq)]A 
\end{eqnarray}
in the interval $(0,T)$ with the periodic boundary conditions $A(q,T)=A(q,0)$ and $B(q,T)=B(q,0)$. Floquet theorem ensures that the quasi-energy spectrum comprises two branches $E(q)= E_{\pm}(q)$, like for the static lattice model discussed in Sec.II.A, with corresponding Floquet states $ \phi_{\pm}(q,t)=(A_{\pm}(q,t), B_{\pm}(q,t))^T$. Note  that, if the dynamics of the system defined by Eqs.(A1) and (A2)  is observed at discretized times $\tau=0,T,2T,...$, from Eq.(A5) and owing to the periodicity of the functions $A(q,t)$ and $B(q,t)$ it follows that it is equivalent to the dynamics of a static lattice with two minibands whose dispersion relations $E_{\pm}(q)$ are given by the quasi-energies of the periodic system. In fact, after setting $\psi(n,t)=(a_n(t),b_n(t))^T$, an initial wave packet, obtained from the superposition of Bloch-Floquet states with arbitrary spectra $F_{\pm}(q)$, evolves in time according to the relation
\begin{eqnarray}
\psi(n,t) & = &  \int_{0}^{2 \pi}dq F_+(q) \phi_+(q,t) \exp[iqn-iE_{+}(q)t] \nonumber \\
& + & \int_{0}^{2 \pi}dq F_-(q) \phi_-(q,t) \exp[iqn-iE_{-}(q)t]. \;\;\;\;\;
\end{eqnarray}
If the evolution of the wave packet is observed at discretized times $\tau =l T$ with $l=0,1,2,3,...$, since $\phi_{\pm}(q, l T)=\phi_{\pm}(q, 0)$ is independent of $\tau$, from Eq.(A8) it follows that $\psi(n,\tau)$ shows the same evolution as the one of a wave packet in a static dimerized lattice with energy band dispersion given by the quasi-energies $E_{\pm}(q)$ of the time-periodic lattice [compare Eq.(A8) with Eq.(27) given in the text]. Therefore, all the dynamical aspects of the time-periodic system defined by Eqs.(A1) and (A2), including the onset of $\mathcal{PT}$ symmetry breaking and its convective or absolute nature, can be derived from an equivalent static lattice with a band structure given by the quasi-energy band structure of the original time-periodic system. In particular, as discussed in Sec.III the convective or absolute nature of the symmetry breaking will be determined by the imaginary part of the quasi-energies at the dominant saddle point. \par   
The determination of the quasi-energy spectrum $E_{\pm}(q)$ generally requires to resort to a numerical analysis of Eqs.(A6) and (A7). An approximate analytical form the quasi energies can be obtained, however, in the large frequency limit. In fact, assuming $\kappa_{1}, \kappa_2, \kappa_3,g \ll \omega$,  the change of the amplitudes $A$ and $B$ over one oscillation cycle are small, so that in Eqs.(A6) and (A7) we may neglect the derivative terms $(dA/ dt)$, $(dB/dt)$ and replace the functions $F(t)$, $G(t)$, $H(t)$ with their average values over the oscillation cycle (rotating-wave approximation), namely one can set
\begin{eqnarray}
E(q) A & \simeq  & -\kappa_3[ \langle H \rangle \exp(iq)+\langle H^* \rangle  \exp(-iq)]A \nonumber \\
& + & igA-[\kappa_1 \langle  F \rangle + \kappa_2 \langle G \rangle \exp(-iq)]B \\
E (q) B &  \simeq  & -\kappa_3[\langle H \rangle \exp(iq)+\langle H^* \rangle  \exp(-iq)]B \nonumber \\
& - & igB-[\kappa_1 \langle F^* \rangle + \kappa_2 \langle G^* \rangle  \exp(iq)] A . \; \; \; \; 
\end{eqnarray}
where $\langle ... \rangle$ denotes the time average. Using the identity of Bessel functions $\exp(i \Gamma \sin x)=\sum_{n=-\infty}^{\infty}J_n(\Gamma) \exp(inx)$, from Eq.(A4) one readily obtains
\begin{equation}
\langle F \rangle = \langle G \rangle  =J_M(\Gamma), \;\; \langle H \rangle =J_0(2 \Gamma \cos \phi) \exp(i \varphi) \;\;\;\;
\end{equation}
where $\varphi=M(\pi+2 \phi)$.  A comparison of Eqs.(A9),(A10) with Eq.(5) given in the text shows that, in the large modulation limit, the ac-dc driven lattice described by Eqs.(A1) and (A2) effectively describes the static Rice-Mele lattice [Eqs.(1) and (2) given in the text], where the effective hopping rate $\kappa$, $\sigma$ and $\rho$ are given by Eqs.(19-21) and the phase $\varphi$ by Eq.(18).

\section{Determination of the saddle point for the extended Rice-Mele lattice model}
In this Appendix we calculate the saddle points $q_s$, i.e. the roots $q_s$ of Eq.(29) given in the text, which determine the convective or absolute nature of the $\mathcal{PT}$ symmetry breaking for the non-Hermitian Rice-Mele Hamiltonian (3). To this aim, it is worth introducing the variables $X_s= \cos q_s$ and $Y_s=\sin q_s$. After some algebra, from Eq.(29) it follows that  $X_s$ and $Y_s$ are the roots (in the complex plane) of following system of algebraic equations
\begin{eqnarray}
X_s^2+Y_s^2 & = & 1 \\
\left( \frac{\kappa \sigma}{2 \rho} \right)^2 Y_s^2 & = & \left( \cos^2 \varphi Y_s^2+ \sin^2 \varphi X_s^2-2 \cos \varphi \sin \varphi X_s Y_s\right) \;\;\; \nonumber \\
& \times & (-\epsilon^2+2 \kappa \sigma +2 \kappa \sigma X_s).
\end{eqnarray}
To simplify the analysis, let us consider the case where the gain/loss parameter $g$ is larger but close to its threshold value $g_{th}$, so that $g^2-g_{th}^2=\epsilon^2$ is a small quantity. Note that, for $\epsilon \rightarrow 0$, a solution to Eqs.(B1) and (B2)  is $X_s=-1$ and $Y_s=0$, corresponding to $q_s=\pi$, i.e. to the wave number where the most unstable mode arises at the $\mathcal{PT}$ symmetry breaking threshold. For $\epsilon^2>0$, we look for a solution to Eqs.(B1) and (B2) in the form of power series
\begin{eqnarray}
X_s & = & -1+\frac{\alpha^2}{2}-\frac{\alpha^4}{4 !} + ... \\
Y_s & = & -\alpha +\frac{\alpha^3}{3 !} -\frac{\alpha^5}{5!}+ ...  \; ,
\end{eqnarray}
where $\alpha=q_s-\pi$ is a small amplitude of order $\epsilon^ \gamma$ with $\gamma>0$ to be determined.  Note that, at leading order in $\alpha$, the energy $E_{+}(q_s)$ is given by
\begin{equation}
E_+(q_s) \simeq -2 \rho \cos \varphi - 2 \rho \sin \varphi \alpha +\sqrt{\kappa \sigma \alpha^2-\epsilon^2}.
\end{equation}
 Note  also that, with the Ansatz (B3) and (B4), Eq.(B1) is automatically satisfied for any $\alpha$. The small complex amplitude $\alpha$ can be determined by substitution of Eqs.(B3) and (B4) into Eq.(B2) and letting equal the terms of lowest order on the left and right hands of the equations so obtained. Three cases should be distinguished.\\
 \\
(1) $|v_g| \neq \sqrt{\sigma \kappa}$, where $v_g=-2 \rho \cos \varphi$.\\ 
In this case Eq.(B2) is satisfied at leading order for $\alpha \sim \epsilon $ (i.e. $ \gamma=1$), namely one obtains 
\begin{equation}
\alpha^2=\frac{\epsilon^2 v_g^2}{\kappa \sigma (v_g^2-\kappa \sigma)}.
\end{equation}
For $v_g^2 > \sigma \kappa$, the two roots $ \alpha$ of Eq.(B6) are real-valued, and correspondingly the imaginary part of $E_{+}(q_s)$, with $q_s=\pi+\alpha$, vanishes [see Eq.(B5)]. Therefore,  for $v_g^2 > \sigma \kappa$ one has $\psi(n,t) \rightarrow 0$ as $t \rightarrow \infty$ along the ray $n/t=0$, i.e. the $\mathcal{PT}$ symmetry breaking is convective. Conversely, for $v_g^2 < \sigma \kappa$ according to Eq.(B6) the amplitude $\alpha$ is purely imaginary, and correspondingly for one of the two roots  the imaginary part of $E_{+}(q_s)$ is positive according to Eq.(B5). In this case $|\psi(n,t)| \rightarrow \infty $ as $t \rightarrow \infty$ along the ray $n/t=0$, i.e. the $\mathcal{PT}$ symmetry breaking is absolute.\\\
\\
(2) $|v_g| =  \sqrt{\sigma \kappa}$ and $\varphi \neq \pm \pi/2$. In this case one obtains $\alpha \sim \epsilon^{2/3}$, i.e. $\gamma=2/3$, and $\alpha$ satisfies the cubic equation
\begin{equation}
\alpha^3= - \frac{\sin \varphi \epsilon^2}{2 \kappa \sigma \cos \varphi}.
\end{equation}
Two of the three roots of such an equation are complex-valued, and correspondingly one can readily shown from Eq.(B5) that a positive imaginary part for the energy $E_+(q_s)$ arises from one of the two complex roots. In fact, since $\epsilon^2$ is of higher order than $\alpha^2$ and $2 \rho \cos \varphi = \sqrt{\kappa \sigma}$, from Eq.(B5) one has 
${\rm Im} \{ E_+(q_s)\} \simeq   2 v_g  {\rm  Im} (\alpha)$.
Therefore in this case the $\mathcal{PT}$ symmetry breaking is absolute.\\
\\
(3) $|v_g| =  \sqrt{\sigma \kappa}$ and $\varphi = \pm \pi/2$. In this case one has $\alpha \sim \epsilon^{1/2}$, i.e. $\gamma=1/2$, and $\alpha$ satisfies the quartic equation
\begin{equation}
\alpha^4=-\frac{\epsilon^2}{\kappa \sigma}.
\end{equation}
 The four roots of such equation are complex-valued, two with positive and two with negative imaginary parts. Correspondingly, like in the previous case a positive imaginary part for the energy $E_+(q_s)$ does appear because ${\rm Im} \{ E_+(q_s)\} \simeq   2 v_g  {\rm  Im} (\alpha)$. Therefore the $\mathcal{PT}$ symmetry breaking is absolute like in the previous case.


\begin{thebibliography}{99}
\bibitem{note}
The Hamiltonian (3) reduces to the Rice-Mele model for $\rho=0$ and after the substitution $g \rightarrow ig$, which makes the Hamiltonian Hermitian.
\end{thebibliography}
\end{document}